# Preparation of YBCO superconducting thick films On MgO substrates by modified melt growth process

Y.B. Zhu*, J.W.Xiong, , S. F. Wang, , Y. L. Zhou, Q. Zhang, S. Y. Dai, Z. H. Chen,

H.B.Lu, and G. Z.Yang

Laboratory of Optical Physics, Institute of Physics and Center for Condensed Matter Physics,
Chinese Academy of Sciences, Beijing 100080, China

---

*Corresponding author.Tel.:+86-10-8264-9299.

E-mail address: zyb@aphy.iphy.ac.cn

**ABSTRACT.**

Superconducting thick films (200μm ~300μm) of YBCO have been fabricated successfully on MgO substrates by a new approach. The precursor powders ($YBa_2Cu_3O_{7-\delta}$(Y123) + 0.1mol$Y_2BaCuO_5$(Y211)) are placed between two MgO (10mm×10mm) slices to form sandwich structure. The YBCO thick films have been obtained from the precursors by modified melt growth process. Resistance measurements of YBCO thick films show Tcon of 87.3K and ΔT of 4.5K. Estimates using hysteresis loops and the Bean model give a value of $2.78\times10^3$ A/cm$^2$ (77K, 0T) for the critical current density. The observations of scanning electron microscopy (SEM) and x-ray diffraction patterns (XRD) show the supercoducting Y123 phase matrix containing discrete Y211 inclusions.

**1. INTRODUCTION**

It is well known that thick films of high-temperature superconducting YBCO widely be used in microwave and power engineering applications. Specially, the application of superconducting thick films to power technologies may offer significant advantages over the use of bulks and thin films. Numerous fabrication techniques of the superconducting YBCO thick films have been investigated using various methods, such as screen-printing, doctor-blade, electrophoresis, etc. [1-5]. All methods allow for the inexpensive production of large-area coatings and complex shape of substrates. The development of thick films of YBCO on YSZ substrates has been studied intensively [6-13]. It has been proved that the key step to improve the superconducting properties of YBCO thick films was heat treatment above the peritectic temperature. In their studies, the MgO substrates were unsuitable for the production of superconducting thick films due to heating above the peritectic temperature [11-13]. Recently, we succeeded in obtaining superconducting thick films on MgO substrates by a new approach. The sinter temperature is 1030°C that exceeded the peritectic temperature.

In this report, we describe the new approach in detail, and present the properties of the superconducting YBCO thick films fabricated by this method.

## 2. EXPERIMENTS

### 2.1. Preparation of precursor powders

The Y123 and Y211 powders were sintered in a square furnace in air by solid-state reaction. The purity of raw materials were $Y_2O_3$ (99.9%), $BaCO_3$ (99.9%), and CuO (99.9%), respectively. The above materials were weighed and mixed to obtain Y123 (Y211) powders with the atoms ratio of 1:2:3 (2:1:1). Firstly, the Y123 and Y211 powders were sintered from room temperature to 900°C and kept at that temperature for 12 hours. Secondly, the temperature was reduced to 30°C. Thirdly, the precursors were ground into thin powders. The calcination temperature being 920°C ( hold for 24 hours) and 940°C ( hold for 48 hours), respectively, the above procedure was repeated twice. The black Y123 powder and the green Y211 powder were obtained. After mixing Y123 powder and 0.1mol Y211 powder and sintering at 950°C for 48 hours, the precursors were ground into fine Y123+Y211 powders. The particle size of the powder ranged from 1μm to 5μm.

**2.2. Fabrication of superconducting YBCO thick films**

The precursor powder without any viscosity solvent was bestrewed direct on the cleaned MgO (10mm×10mm) substrates by a mesh. Then another cleaned MgO (10mm×10mm) slice was placed carefully on the powders to form a sandwich structure. The sample was put into a square furnace and heatd up to 1030°C in 4 hours, kept at 1030°C for 15 minutes, then slowed down the temperature to room temperature in 10 hours. After the heat treatment, the upper MgO slice was peeled off and YBCO thick films on MgO can be obtained. The experimental process was reproducible for producing the good superconducting YBCO thick films.

The surface morphology of the superconducting YBCO thick films was illustrated by scanning electron microscopy (SEM). The crystalline alignment of the thick films was examined with the x-ray diffraction analysis (XRD). Zero-resistance transition was measured by standard four-probe method. The critical current density was derived from magnetic measurement.

**3. RESULTS AND DISCUSSION**

The superconducting YBCO thick films obtained in this work are black, shining, flat and compact. The upper MgO slice makes the films very smooth. The powder without any viscosity solvent permits the particles of the powder to self-organize freely, as a result that the weak link of grain boundaries was reduced greatly. Since the samples are sintered above peritectic temperature, the superconducting properties are improved significantly. Moreover, the diffusion between the interfaces is avoided effectively, as the peritectic temperature is kept in a short time. These are the main reasons that the thick films achieved good superconductivity.

Analysis of the films using XRD (Fig. 1) shows mostly orientated peaks of Y123 phase and a few small peaks of Y211 phase. The values of the relative intensity of (003), (005), (006) peaks in Y123 phase are obviously greater than that of other peaks presented in the XRD patterns, indicating most crystalline grains of the thick films are highly textured in the 00*l* direction.

The images of surface morphology by scanning electron microscopy (Fig. 2) illustrates well grain-aligned boundaries of Y123 matrix and Y211 particles trapped within Y123 domains with a random mode. That is corresponded with one model of the distribution of trapped Y211 particles in Y123 domains described in reference [13].

Standard four-probe method is used to measure the resistance of YBCO thick films. It shows in figure 3 that the onset transition temperature is 87.3K and the offset transition temperature is about 82.8K. These results can be compared with the superconductivity of YBCO/MgO thick films prepared by electrophoresis [14].

Magnetization measurement is carried out to obtain the critical current density. The external magnetic field is vertical to the surface of the thick films of YBCO. Using hysteresis loops and the Bean model, the critical current density versus the external magnetic field at 77K is shown in figure 4. Zero field transport critical current density is up to $2.78\times10^3$ A/cm$^2$ at 77K, which is similar to the situation in reference [15].

## 4. CONCLUSIONS

We develop a low cost, simple equipment, higher growth rate approach to grow the thick films of superconducting YBCO by a modified melt growth process. The superconducting properties of the thick films are examined by common measurement methods. The characterization of the thick films are in high quality to be compared with that of the thick films ever reported. Since the samples are sintered above peritectic temperature, the superconducting properties are improved significantly. Moreover, the diffusion between the interfaces is avoided effectively, as the peritectic temperature is kept in a short time. These are the main reasons that the thick films achieved good superconductivity.

## ACKNOWLEDGEMENTS

This work was supported by a grant of the State Key Program of China and by National Natural Science Foundation of China.

**Figure Captions:**

Figure.1 XRD patterns for the YBCO/MgO film.

Figure.2 SEM image for the surface of the YBCO/MgO film

Figure.3 Resistance versus temperature for the YBCO/MgO film. The inset is a magnified view for temperatures near $T_{c.}$

Figure.4 The critical current density of the YBCO/MgO thick fillm versus the external magnetic field at 77K.

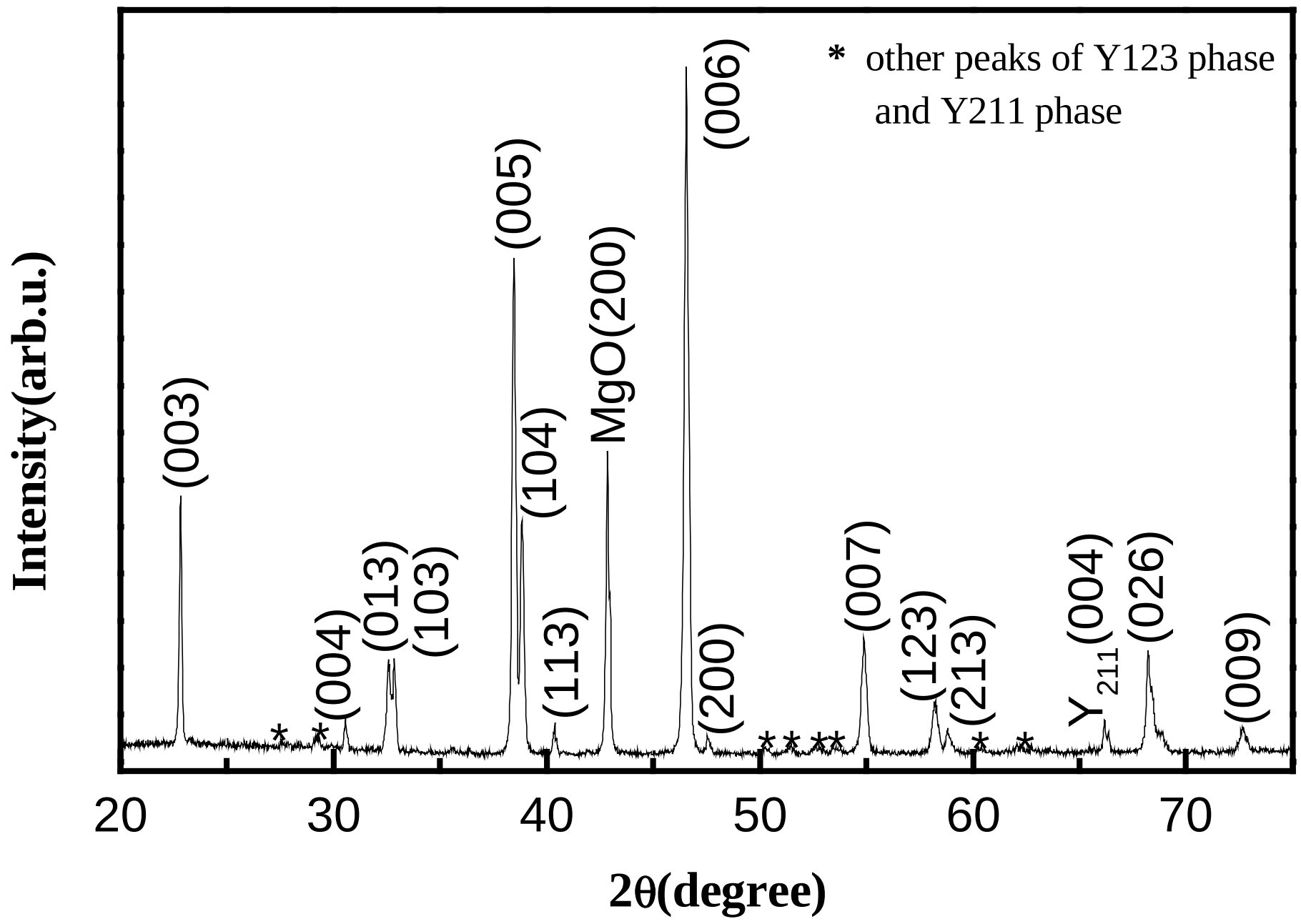

**Figure. 1  Y.B. Zhu et al.**

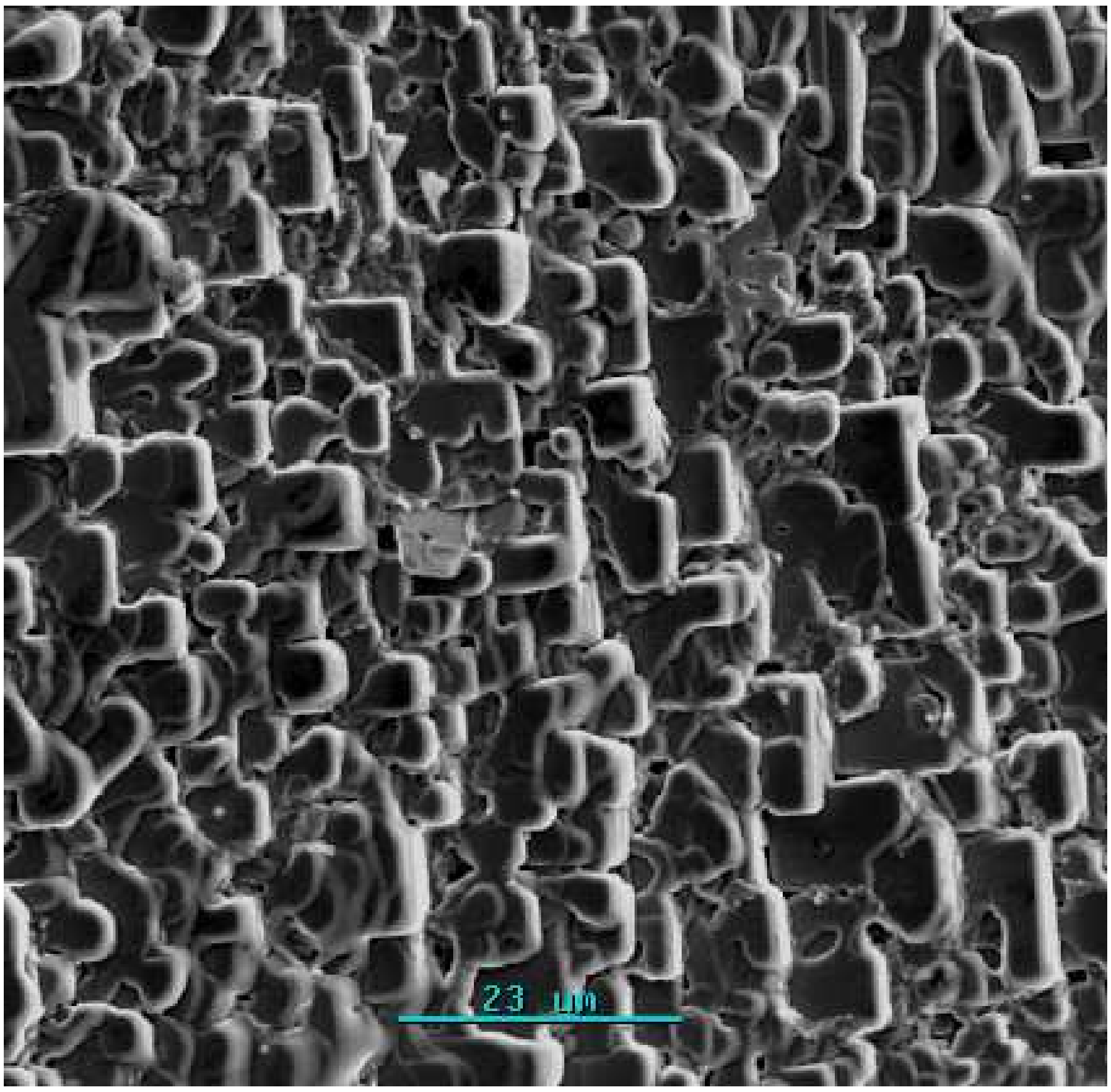

**Figure. 2 Y.B. Zhu et al.**

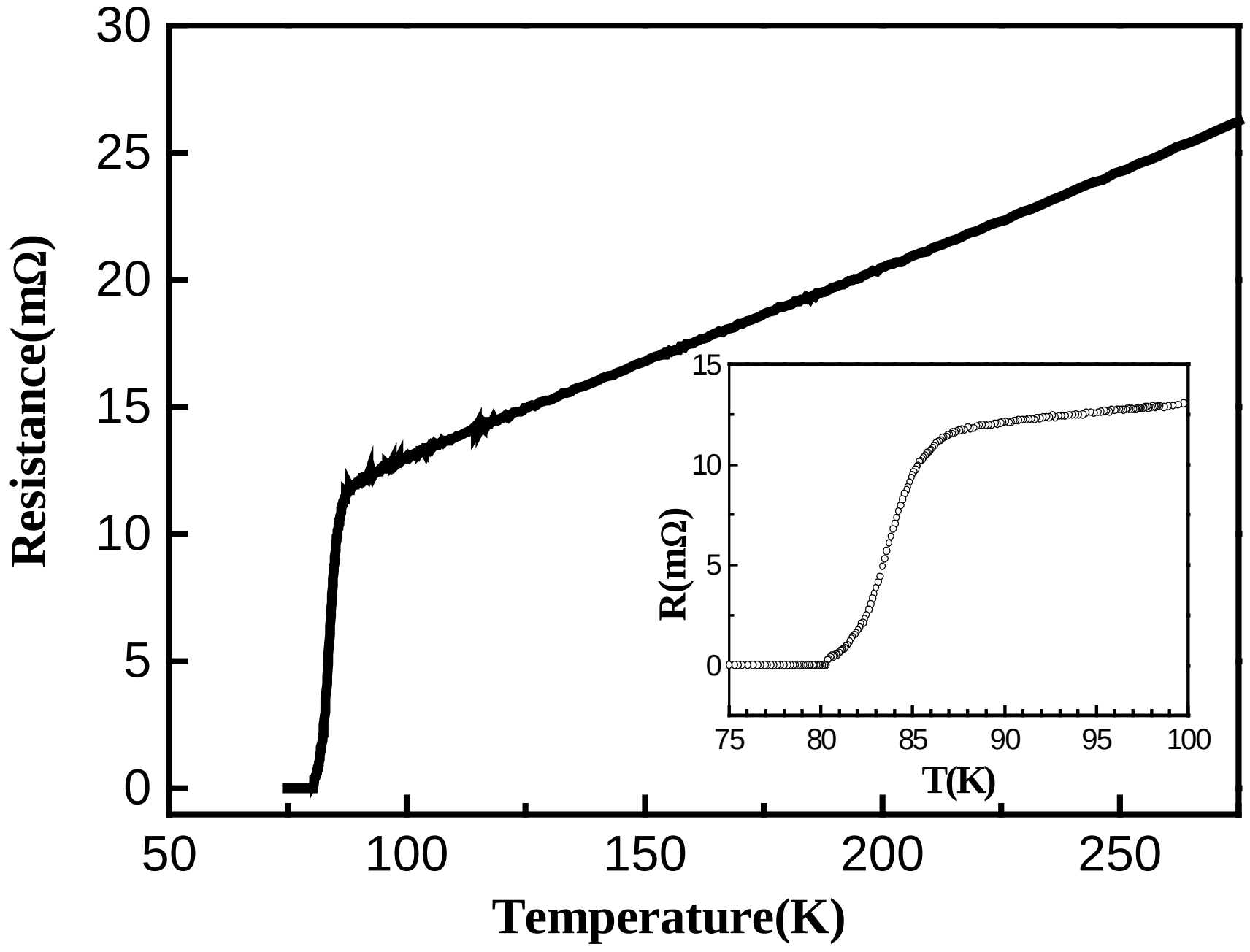

**Figure. 3    Y.B. Zhu et al.**

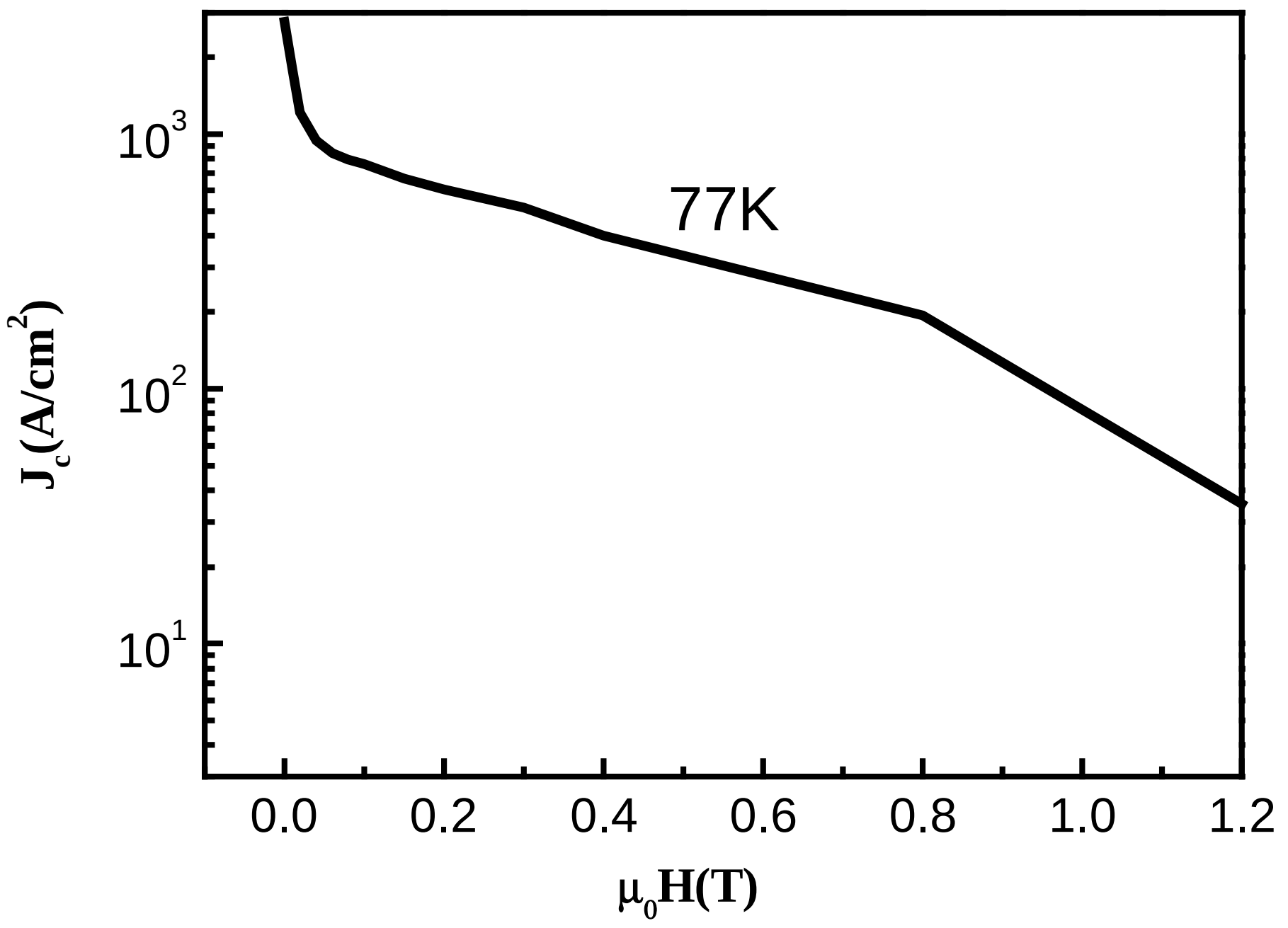

**Figure.4    Y.B. Zhu et al.**